\documentclass[11pt]{article}
\usepackage[utf8]{inputenc}
\usepackage[T2A]{fontenc}               
\usepackage[russian,english]{babel}                 
\usepackage{epsfig,graphics,color}
\author{\large \bf  Zafar U. Usubov\footnote      
       {On leave of absence from Institute of Physics, Baku, Azerbaijan}
\\
\\Joint Institute for Nuclear Research,
\\ Dubna, Russia}          

\title { Electromagnetic calorimeter simulation for future   \\      
         $\mu \to e$ conversion experiments }                  

\begin{document}
\maketitle
{
\vskip 0.5cm
\hskip 6.5cm {\bf \large { Abstract}}
\vskip 0.5cm

{\large {
We examine three dense  high-Z scintillating crystals 
for the $\mu \to e$
conversion experiment using  the GEANT4 simulation toolkit. The
full energy deposition, albedo, and longitudinal and lateral energy leakages
for all crystal  assemblies are studied.
The influence of the crystal depth on the energy deposition in the calorimeter is  studied.

\Large {
\section{Introduction}
\vskip -0.8cm
$ $

Flavor changing by all neutral current interactions 
is strongly suppressed in the Standard Model(SM)
and has not yet been observed.
The present values of the neutrino oscillations parameters lead  to 
Br($\mu \to e\gamma$) $\le$ 10$^{-54}$ in the SM.
Review of the modern theoretical motivations for lepton flavor violation, 
data from current experimental
bounds and expected improvements astonishingly collected 
by Marciano, Mori and Roney\cite{marc}.
Many scenarios beyond the SM (supersymmetry, extra dimensions,
little Higgs, quark compositeness)  naturally allow  and 
predict  some experimentally observable level for processes with charged lepton
flavor violation(CLFV).
 The $\mu \to e$ conversion experiments 
are 10 to 100 times more sensitive to new physics
than CLFV searches in other channels.                
        To date
the upper limit on the $\mu^{-} \to e^{-}\gamma$  branching ratio 
is $2.4\cdot10^{-12}$ at 90\% CL\cite{adam}.

The aim of the $\mu \to e$ conversion experiments is to search 
for the coherent conversion of muons
to electrons in the field of a nucleus.  
Many of the ideas of such experiments
are based on a concept that was first proposed by Djilkibaev and Lobashev for
the MELC experiment\cite{djil}, then was developed in MECO\cite{meco} project
and now formulated in the Mu2e proposal\cite{mu2e}.

At the first step high intense proton beam is directed to the production target.   
Then pions focused with the graded field  produce  high-intensity  low-energy
($p_{\mu}^{-} < 100\,$MeV) muon beam which serves to create  muonic atoms 
in thin targets.
A muon  in the orbit could then ordinarily   decay or weak capture  on the nucleus. 
For muonic aluminum these processes occur with a probability 
$\sim$0.4 and $\sim$0.6, respectively.

Muon in the orbit can  be
converted to an electron through some new CLFV interaction with the 
nucleus. The conversion-to-capture  ratio 
$$ R_{\mu e} = {{\Gamma(\mu^{-}N \to e^{-}N)} 
\over {\Gamma(\mu^{-}N \to {\nu}_{\mu}N^{'})}}
$$
\begin{figure}[h!]                            
{\vskip -4.0cm}
{\hskip  0.8cm}  {\epsfxsize  6.0 truein \epsfbox{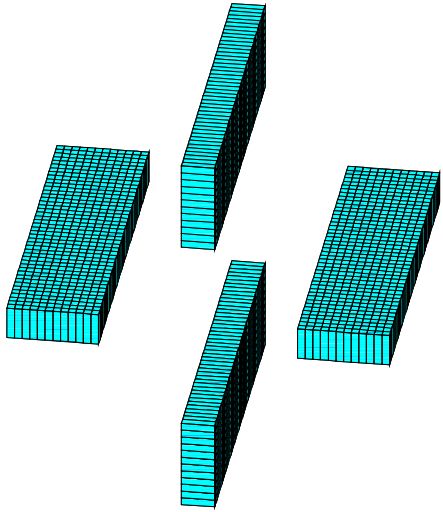}}
{\vskip -4.0cm}
\caption[]{Trigger electromagnetic calorimeter for the Mu2e experiment.}
\label{Norm}
\end{figure}

is $< 6.1(7)\cdot 10^{-13}$ at 90\% CL  for
titanium(gold) nuclei, as obtained in the  SINDRUM II\cite{sin2} experiment.
For a nominal two-year run (2$\cdot 10^{7}$s) the  Mu2e experiment     
could discover  about 40 signal events on a background of less              
than 0.5 events\cite{kuts}.
The expected limit on $R_{\mu e}$  is $6\cdot10^{-17}$ 
at 90\% CL. This corresponds to the effective mass about 10$^4\,$TeV.
\begin{table}[]
\large {
\begin{center}
{\color{black}        
\begin{tabular}{||c||c||c||c||c||}     \hline \hline
  Crystal                      & NaI(Tl) & BGO    &LYSO(Ce) & PWO \\ \hline \hline
  Density\,(g/cm$^3$)           & 3.67    & 7.13   & 7.40    & 8.30 \\ \hline 
 Melting  Point  (${^0}$C)      & 651   &1050  &2050  & 1123 \\ \hline        
 Radiation  Length\,(cm)       &2.59   &1.12  &1.14  & 0.89 \\ \hline            
Moli\`ere  Radius  (cm)         & 4.13  &2.23  &2.07  & 2.00 \\ \hline           
 Interaction  Length\,(cm)     & 42.9  &22.8  &20.9  & 20.7 \\ \hline            
 Refractive  Index             & 1.85  &2.15  &1.82  & 2.20 \\ \hline           
 Hygroscopicity                & Yes   & No   & No   &  No  \\ \hline            
 Luminescence  (nm)(at  peak)  & 410   & 480  & 402  & 425(420) \\ \hline            
 Decay  Time  (ns)             & 245   & 300  &  40  & 30(10)\\ \hline            
 Light  Yield$(\%)$              & 100   & 21   &  85  & 0.3(0.077) \\ \hline            
 d(LY)/dT($\%/{^0}C$)            &-0.2   &-0.9  &-0.2  &-2.5  \\ \hline \hline            
\end {tabular}
\caption{Useful characteristics\cite{maoo}  of dense crystals as a Mu2e calorimeter material.
The values correspond to  the slow or fast(in parentheses)  scintilation component.}
}
\end{center}
}
\end{table}

In this paper, we explore three dense crystals for the Mu2e trigger       
calorimeter using the GEANT4\cite{gean} simulation toolkit.
The next section gives a brief description of the Mu2e experiment setup  and
the choice of dense crystals for the calorimeter. Section~3 gives
our GEANT4 simulation strategy. In this section we present the 
full energy deposition in the calorimeter, albedo, longitudinal 
and lateral energy leakages, and  the influence of tyvek wrapping on the crystal.
We end with the conclusions in Section~4.     

\section{Experiment setup and crystal choice}              
\vskip -0.8cm
$ $

The first stage of the  Mu2e experiment  
will be performed at  Fermilab using re-bunched and          
slow-extracted 8 GeV proton beam from the booster. 
The Mu2e setup consists of a production solenoid with a production
\begin{figure}[h!]                            
{\vskip  1.5cm}
{\hskip  0.0cm}  {\hskip 3.0cm \epsfxsize  4.0 truein \epsfbox{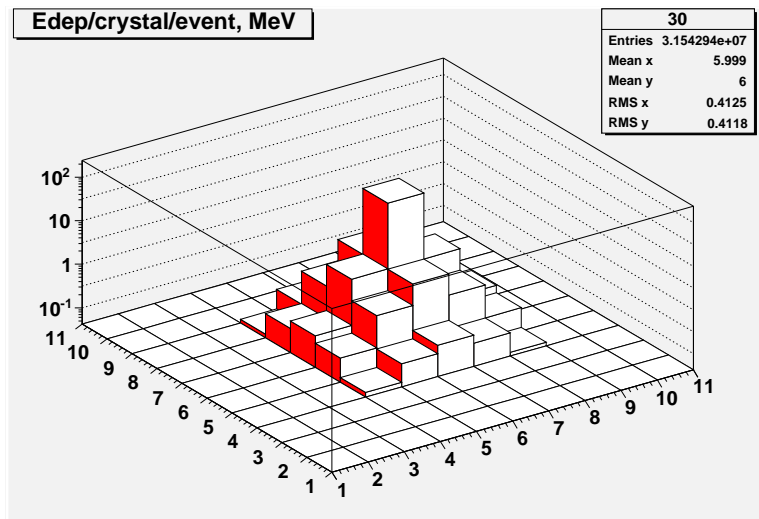}}
{\vskip -0.0cm}
\caption[]{Lego plot for energy deposition in LYSO(Ce) crystals. The electrons with E=105 MeV penetrate the calorimeter
perpendicularly.}
\label{Norm}
\end{figure}

\begin{figure}[h!]                            
{\vskip  0.5cm}
{\hskip  0.7cm}  {\epsfxsize  6.0 truein \epsfbox{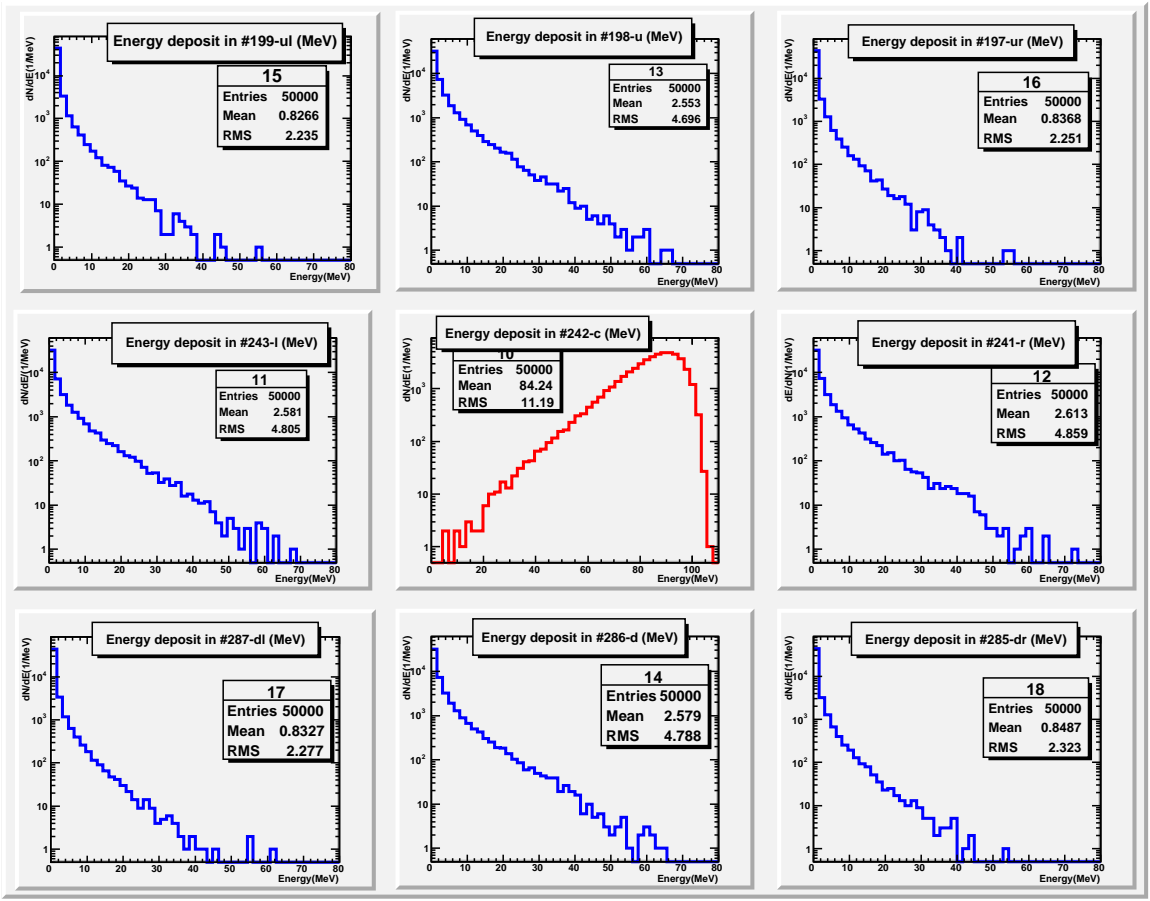}}
{\vskip -0.0cm}
\caption[]{Energy deposition in the 
central 3$\times$3 matrix of the LYSO(Ce) crystals. The electrons with E=105 MeV penetrate the calorimeter
perpendicularly.}
\label{Norm}
\end{figure}
              
target, a transport solenoid with collimators, and a detector solenoid
with a stopping target, tracker, and electromagnetic calorimeter.
The detector solenoid will provide a  highly uniform magnetic field of 1.0 T
in tracker and calorimeter area.

The detector of    interest  is the electromagnetic  calorimeter
designed as a trigger system,    which is also  
used for measuring (in addition to tracking system)  the energy and 
direction  of the conversion  electron emitted at the energy $\sim$105.1~MeV   
(for aluminum stopping  target). 
Dense crystals with a short radiation length,  small  Moli\`ere radius,
and short decay time can offer good space and time resolution for the calorimeter 
and allow    reduction of the  background in the energy window corresponding to the
conversion electron.

Crystal scintillators have long been used in nuclear and high-energy physics\cite{webe}.
Rutherford\footnote{The Nobel Prize in Chemistry 1908.} 
used  ZnS in
his alpha-particle scattering study\cite {reth}, in 1948 Hofstadter\footnote{The Nobel Prize in 
Physics 1961, was divided with R.L.\,M\"ossbauer.}
first demonstrated NaI(Tl) as a general-purpose detector for photon
spectroscopy\cite{hofs}, CsI(Tl) crystals are now used
by BELLE, BABAR, BES~III, and
PbWO$_4$ by CMS and ALICE at the LHC.

Potentially desirable crystals to fully contain electromagnetic showers 
at the Mu2e  experiment energy are lead tungstate 
PbWO$_4$(PWO), bismuth germanate Bi$_4$Ge$_3$O$_{12}$(BGO),     
cerium-doped lutetium oxyorthosilicate 
Lu$_2$SiO$_5$:Ce(LSO(Ce)), and cerium-doped lutetium yttrium oxyorthosilicate
Lu$_{2(1-x)}$Y$_{2x}$SiO$_5$:Ce(LYSO(Ce)).           
The properties of these crystals are compared with  NaI(Tl) in Table~1\cite{maoo}.
This simulation is based on the LYSO crystals which were
activated  with 0.4\% mol of cerium  and $x=0.1$.

In the past years oxyorthosilicates was mainly used for
medical imaging, namely  in the positron emmision tomography
and mammography scanners(small-size good-quality
cryctals)\cite{eijk}.
\begin{table}[]
\large {
\begin{center}
{\color{black}        
\begin{tabular}{||c||c||c||}     \hline \hline
 Crystal  &$E_{dep}^8/E_{dep}^{9}$ &$E_{dep}^9/E_{dep}^{all}$  \\ \hline \hline     
 BGO      &  0.144          &   0.968            \\
 PWO      &  0.133          &   0.974          \\
 LYSO(Ce) &  0.140          &   0.969          \\ \hline \hline
\end {tabular}
\caption{ The relative energy deposition in the central 3$\times$3 matrix of the crystals (see the text).}
}
\end{center}
}
\end{table}

\begin{table}[]
\large {
\begin{center}
{\color{black}        
\begin{tabular}{||c||c||c||c||c||}     \hline \hline
          &$E_{dep}(MeV)       $& Albedo (MeV)    & Longitudinal(MeV)    & Lateral(MeV)   \\                   
 Crystal  &Naked\quad Wrapped&Naked\quad Wrapped&Naked\quad Wrapped &Naked\quad Wrapped\\ \hline \hline     
          &101.1\qquad 101.2   &1.139\qquad 1.139 &3.184\qquad 3.169 &1.917\qquad 1.735\\                   
 BGO      &5.932\qquad 6.031   &1.606\qquad 1.662 &4.403\qquad 4.361 &1.323\qquad 1.406 \\
          &50000\qquad 50000   &39611\qquad 39404 &46353\qquad 47155 & 370 \qquad 385   \\ \hline
          &102.5\qquad 102.6   &1.150\qquad 1.142 &2.843\qquad 2.824 &1.976\qquad 2.077\\                   
 PWO      &4.104\qquad 4.138   &1.512\qquad 1.518 &3.513\qquad 3.410 &1.457\qquad 1.532\\
          &50000\qquad 50000   &45896\qquad 45837 &24042\qquad 24798 & 175 \qquad 188  \\ \hline
          &101.0\qquad 101.1   &0.990\qquad 0.982 &3.131\qquad 3.171 &1.981\qquad 2.079\\                   
 LYSO(Ce) &6.080\qquad 6.286   &1.397\qquad 1.449 &4.543\qquad 4.642 &1.770\qquad 1.731\\
          &50000 \qquad 50000  &40554\qquad 40706 &50003\qquad 50396 & 341 \qquad  413 \\ \hline \hline
\end {tabular}
\caption{The energy deposition, albedo, and lonitudinal and lateral energy leakages for dense crystals with and
without tyvek wrapping (see the text).}
}
\end{center}
}
\end{table}

The proposed calorimeter\cite{mu2e} consists  of four vanes of detectors(Fig.~1).
Each vane considered here comprises  $13\times 3\times 3$  cm$^3$ crystals with the total
size  $13\times 36 \times 132$ cm$^3$. Thus, each calorimeter vane consists  of 528
cells. The interior faces of the vanes are  39.0 cm  away       
from the solenoid axis.
The depth of the calorimeter (13 cm)  corresponds to 11.6, 11.4, and 14.6
radiation lengths for the BGO, LYSO(Ce), and PWO crystals, respectively.
The electrons  are incident on  the $36\times 132$ cm$^2$ side of the calorimeter
at a mean angle of 
55$^0$. 
\newpage
\begin{figure}[h!]                            
{\vskip -0.0cm}
{\hskip  3.8cm}  {\epsfxsize  3.5 truein \epsfbox{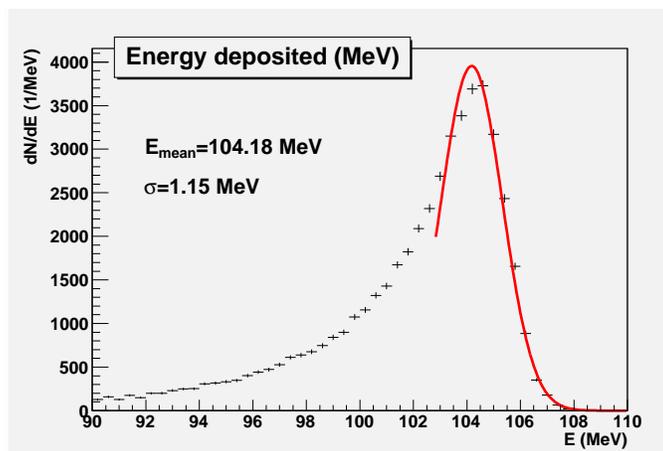}}
{\vskip -0.0cm}
\caption[]{Energy deposition in the calorimeter with the LYCO(Ce) crystals. The electrons with  E=105  MeV
penetrate  the calorimeter  perpendicularly.}
\label{Norm}
\end{figure}

\section{ Calorimeter simulation and results}
\vskip -0.8cm
$ $

The crystal calorimeter simulation is performed with the 
GEANT4 toolkit  using the     
low-energy physics list.
   The minimal tracking step was set to 10$\,\mu$m, which corresponds to the energy
cuts of $\sim$3.2~keV for photons  
and $\sim$51.9(50.8)~keV for electrons(positrons)
for the LYSO(Ce) crystals.

We have studied three parts of shower leakage\cite{wigm}:
longitudinal, lateral, and albedo - the backward leakage through the front face
of the detector.
For each cell,   to the energy deposition  we added a random energy according to the
Gaussian distribution with  $\sigma$=1.0~MeV\cite{luca} and E$_{mean}$=0.0~MeV.
 The same procedure was
applied to the  albedo and the tracks leaving the calorimeter in the longitudinal
and lateral directions.  

In this analysis the optical processes (such as scintillation, refraction,
transparency, propagation of light through the crystals towards the light 
sensors,  conversion of the photons into electronic signals)
was not considered.

Each analysis was based on 50000 generated events.

The lego plot in Fig.~2 presents           
the energy deposition  in the LYSO(Ce) crystals 
of the calorimeter vane. The electron beam  with E=105 MeV 
perpendicularly hits at
the center of the crystal in the middle of the vane. It is seen that
the signal is transversely contained in a 5$\times$5 matrix.
A similar situation is observed for the BGO and PWO crystals.
\begin{figure}[h!]                            
{\vskip -0.0cm}
{\hskip  2.1cm}  {\epsfxsize  5.0 truein \epsfbox{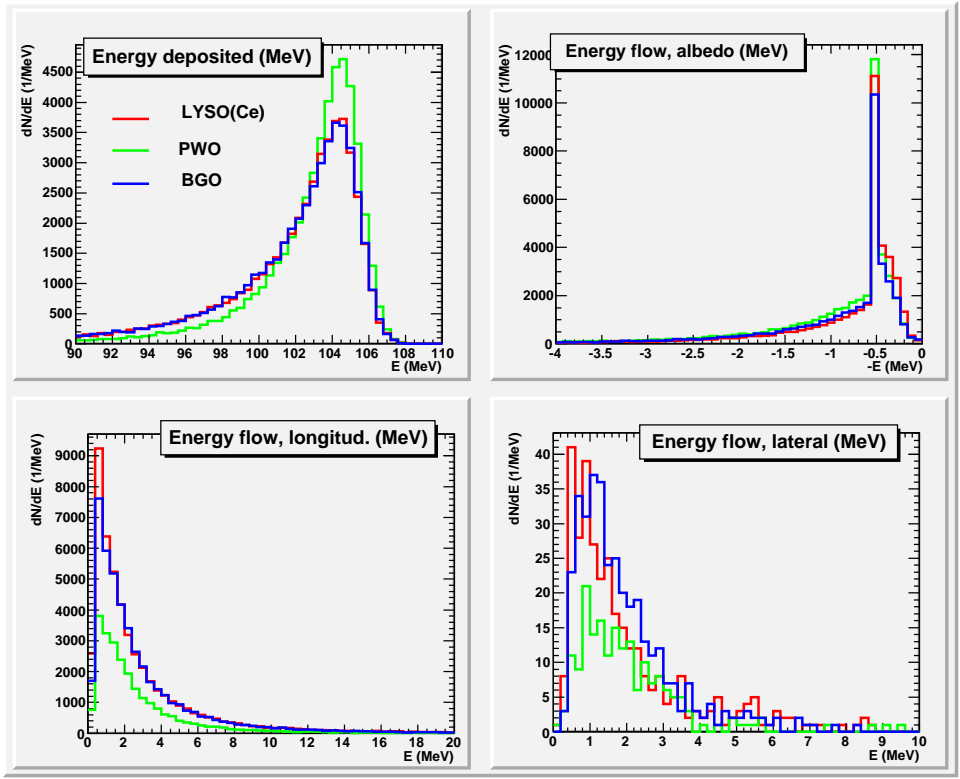}}
{\vskip -0.0cm}
\caption[]{Energy deposition, albedo, and  longitudinal and lateral energy leakages for all
crystal  assemblies. The electrons with E=105 MeV penetrate the calorimeterter in the perpendicular
direction.}
\label{Norm}
\end{figure}

In Fig.~3 we show the energy deposition in all crystals of the            
central $3\times3$ matrix.
Table~2  presents  the ratios
of energy deposition in eight surrounding crystals  to the energy
deposition in all nine crystals in the central $3\times 3$ matrix for 
the BGO, PWO, and LYSO(Ce) crystals.
The ratio of the 
energy deposition in the central $3\times 3$ matrix        
to the total  energy deposition in a vane is also presented in the table.
We note that most of the  105~MeV
electron energy ($>$95\%) is  deposited  in the 
central 3$\times$3 matrix. The shortest Moli\`ere radius of PWO is reflected
in the data.               

In Fig.~4  the total  energy deposition in the calorimeter 
composed of LYSO(Ce) crystals and irradiated by perpendicularly
directed 105 MeV electrons is depicted.
The curve corresponding to the
Gaussian fit is also shown in the figure. The parameters of the fit
are E$_{mean}$ = 104.18$\pm$0.012\,MeV and
$\sigma$=1.15 $\pm$0.01\,MeV.

Figure~5 compares the total  energy deposition, albedo, and longitudinal
and lateral energy leakages for the 
LYSO(Ce), PWO and BGO crystals. The elctron beam with the  energy of 
105 MeV is perpendicularly directed into the central crystal.
\begin{figure}[h!]                            
{\vskip  0.0cm}
{\hskip  1.0cm}  {\hskip 1.0cm \epsfxsize  5.0 truein \epsfbox{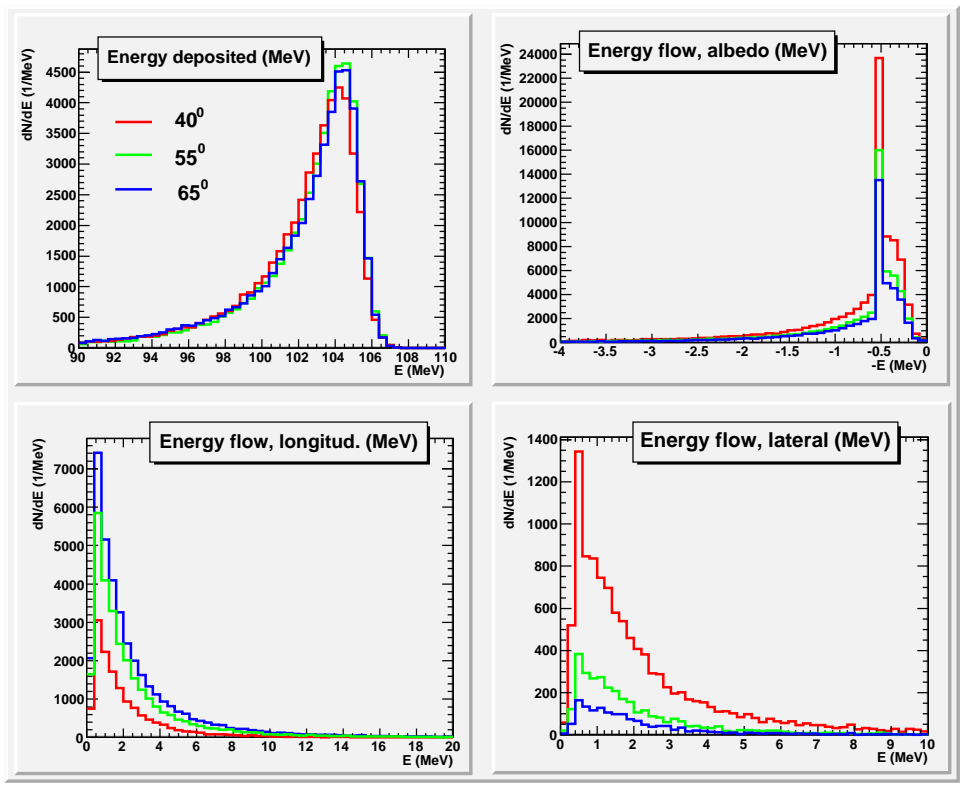}}
{\vskip -0.0cm}
\caption[]{The dependences of the total energy deposition, albedo, and
longitudinal and lateral energy leakages on the electron impact angles for the LYSO(Ce) crystals.}
\label{Norm}
\end{figure}

In the calorimeters the crystals are usualy wrapped  in a reflective
material to protect them from the light of other crystals.
In Table~3 we show the influence of  wrapping of the crystals.           
We modeled  the crystals whose lateral faces were wrapped in 
0.1~mm 
(which corresponds to $\sim$4 mil)  
of tyvek\cite{woj}  consisting of carbon and hydrogen.
The electron beam hits the calorimeter perpendicularly, E=105 MeV.
The left(right) column
in each table cell corresponds to the naked(wrapped) 
crystal of the calorimeter. The data in the rows  show the mean value(top),
root-mean-square(rms) value(middle) and number of entries(lower)
for the energy deposition, albedo, and  
longitudinal and lateral leakages.                        
The data do not show sizable influence of the  wrapping since 
the GEANT4 simulation does not include optical processes.

Note that the lateral energy leakages are almost identical for the 
crystals, although the number of tracks leaving the PWO crystals is
$\sim$2 times smaller than in case of BGO and LYSO(Ce).
\begin{table}[]
\large {
\begin{center}
{\color{black}        
\begin{tabular}{||c||c||c||c||c||}     \hline \hline
 Crystal  &$E_{dep}$(MeV)     & Albedo (MeV)    & Longitudinal(MeV)    & Lateral(MeV)   \\                   
          &$40^0\quad 55^0\quad 65^0$ & $40^0\quad 55^0\quad 65^0$&$40^0\quad 55^0\quad 65^0$&$40^0\quad 55^0\quad 65^0$\\ \hline \hline
          &101.7\,\,102.0\,\,101.7 &1.094\,\,1.040\,\,0.996&2.370\,\,2.761\,\,2.938&2.461\,\,2.134\,\,1.997\\                   
LYSO(Ce)  &4.931\,\,4.761\,\,5.076 &1.878\,\,1.661\,\,1.490&2.945\,\,3.843\,\,4.071&2.614\,\,2.258\,\,1.877\\                   
          &50000\,\,50000\,\,50000 &92618\,\,60112\,\,49816&14221\,\,29338\,\,38488&10965\,\,3210 \,\,1453 \\ \hline \hline     
\end {tabular}
\caption{ The energy deposition, albedo, and lonitudinal and lateral energy leakages for the calorimeter with the LYSO(Ce) crystals
at three different impact
angles of 105 MeV electrons.}               
}
\end{center}
}
\end{table}

\begin{figure}[h!]                            
{\vskip -0.0cm}
{\hskip  2.5cm}  {\epsfxsize  4.5 truein \epsfbox{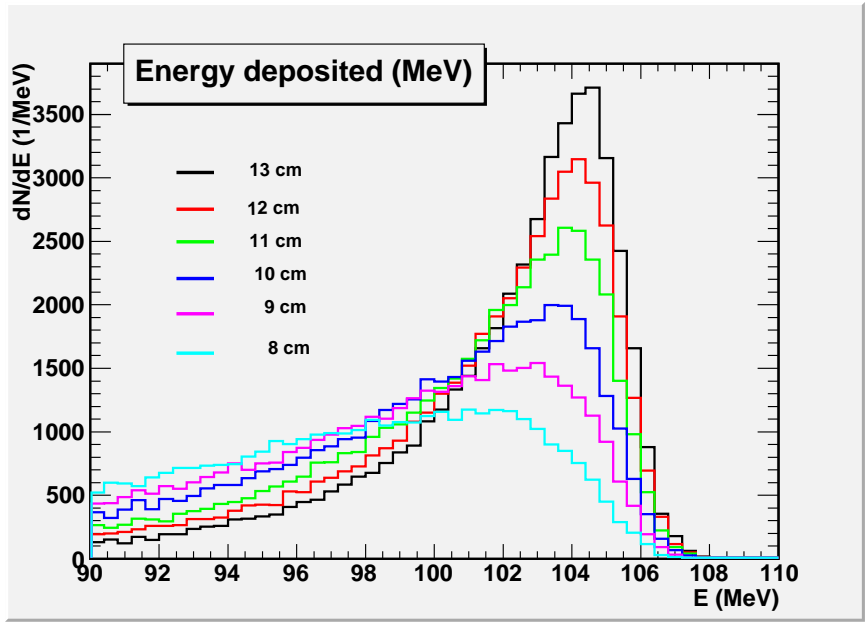}}
{\vskip -0.0cm}
\caption[]{Energy deposition for different LYSO(Ce) crystal depth.                     
The electrons with E=105 MeV penetrate   the calorimeter in the perpendicular
direction.}
\label{Norm}
\end{figure}

The electrons in the magnetic field of the tracking solenoid will move   in   
helical trajectories and impact of the calorimeter
$36\times 132$ cm$^2$ side  with incident
angles ranging from 40$^0$ to 65$^0$, the mean angle being 55$^0$.
In Fig.~6 we display the energy deposition, albedo, and longitudinal and lateral
leakages for these three electron  impact angles relative to the z axis
(the side of the calorimeter in the z-direction is 36 cm) for the LYSO(Ce) crystals.
Table~4 represents
the mean values, rms values  and number of entries for these variables 
in the first, second and third
rows, respectively. Full energy leakages  for these inpact
angles for 105 MeV electrons do   not exceed 6\% for LYSO(Ce).
\begin{figure}[h!]                            
{\vskip -0.0cm}
{\hskip  2.0cm}  {\epsfxsize  5.0 truein \epsfbox{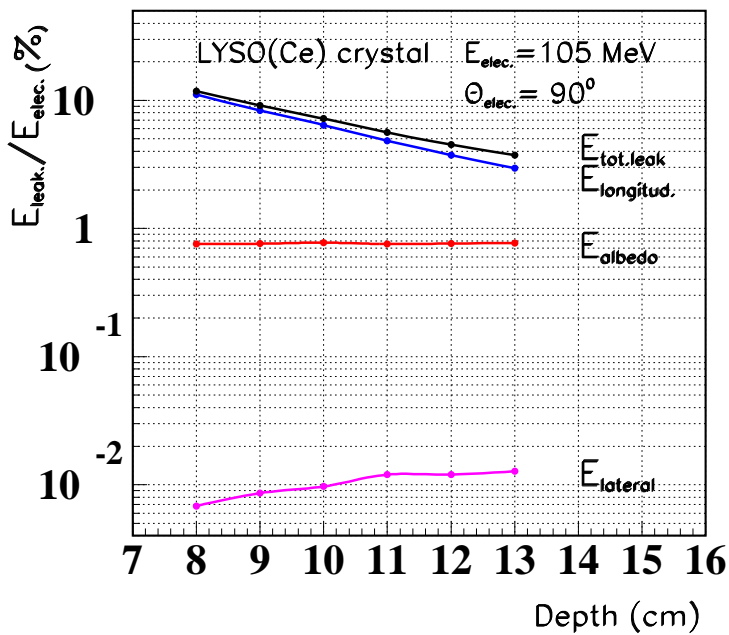}}
{\vskip -1.0cm}
\caption[]{The albedo, longitudinal, lateral, and total energy leakages for the calorimeter with 
LYSO(Ce) crystals of different depth.}
\label{Norm}
\end{figure}

Further, we explore the influence of the crystal depth  on the energy    
deposition and energy leakage  in the calorimeter. In Fig.~7 we demonstrate the energy    
deposition for 105 MeV electrons  and in Fig.~8 the albedo, longitudinal, 
and lateral and total energy leakages for the calorimeter 
with LYSO(Ce) crystals of different depth.

\section{Conclusions}                                            
\vskip -0.8cm
$ $

The ultimate sensitivity and broad new physics motivations
make  $\mu \to e$  coherent conversion a unique channel for                     
charged lepton flavor violating process.
The experiment Mu2e at Fermilab
capable of discovering  the
process or imposing  the upper limit as small as $\sim$2$\cdot$10$^{-17}$,   
would probe new physics at a mass scale up to 
$\sim$10$^{4}$~TeV, much higher than at the  LHC.      

In this paper we performed short comparison study of 
dense  high-Z  PWO, BGO, and 
LYSO(Ce) crystals as candidates for the Mu2e trigger electromagnetic
calorimeter. The study was based on the GEANT4 simulation toolkit.
At this stage, all   crystals showed attractive behavior to  
identify the electron in $\mu \to e$ conversion.

To accurately simulate the experiment, the processes of 
scintillation, refraction,
transparency, propagation of light through the crystals towards the light 
sensors and the conversion of  photons into electronic signals, etc.
must be incorporated in the model.
The calorimeter parameters need to be optimized in the full Mu2e experiment
setup simulation. The LSO/LYSO crystal bench tests are  needed for
various readout devices.

We can expect that 
the high cost of LSO/LYSO crystals ($\sim$15-20 times higher than PWO 
crystals)                                                                      
may be significantly reduced if they are  mass produced for
high-energy physics needs.

}
\

}}
\end{document}